\documentclass[british,english,letterpaper,preprint,superscriptaddress,aps]{revtex4}
\usepackage{graphicx,color,subfigure}
\usepackage{amssymb,amsmath} 
\graphicspath{{figures/}} 

\begin{document}

\title{Spontaneous pattern growth on chocolate surface: simulation and experiments}

\author{Jorge Delgado}
\affiliation{Divisi\'on de Ciencias e Ingenier\'{\i}as, Universidad de Guanajuato, Le\'on, Guanajuato, Mexico}

\author{Claudia Ferreiro-C\'ordova}
\affiliation{Escuela de Ingenier\'ia y Ciencias, Campus Quer\'etaro, Tec de Monterrey, Santiago de Quer\'etaro, Quer\'etaro, Mexico}

\author{Alejandro Gil-Villegas}
\affiliation{Divisi\'on de Ciencias e Ingenier\'{\i}as, Universidad de Guanajuato, Le\'on, Guanajuato, Mexico}

\date{\today}

\begin{abstract}

The natural variation of temperature at ambient conditions produces spontaneous patterns on the surface of chocolate, which result from fat bloom. These patterns are peculiar because of their shape and cannot be obtained by controlled temperature conditions. The formation of these spontaneous grains on the surface of chocolate is studied on experimental and theoretical grounds.Three different kinds of experiments were conducted: observation of formed patterns in time, atomic force microscopy of the initial events on the grain formation and rheology of the melted chocolate. The patterns observed in our experiments follow the trends described by the Avrami model, which considers that is possible to define a characteristic time scale that governs the growth of grains starting from germ nuclei. Through computer simulations, in the NVT ensemble using a coarse-grained model of triacylglycerides molecules, we studied the process of nucleation that starts the pattern growth and that is consistent with the Avrami model.  
\end{abstract}

\maketitle

\section{Introduction}

\begin{figure}[htb]
\begin{center}
\includegraphics[width=0.9\linewidth]{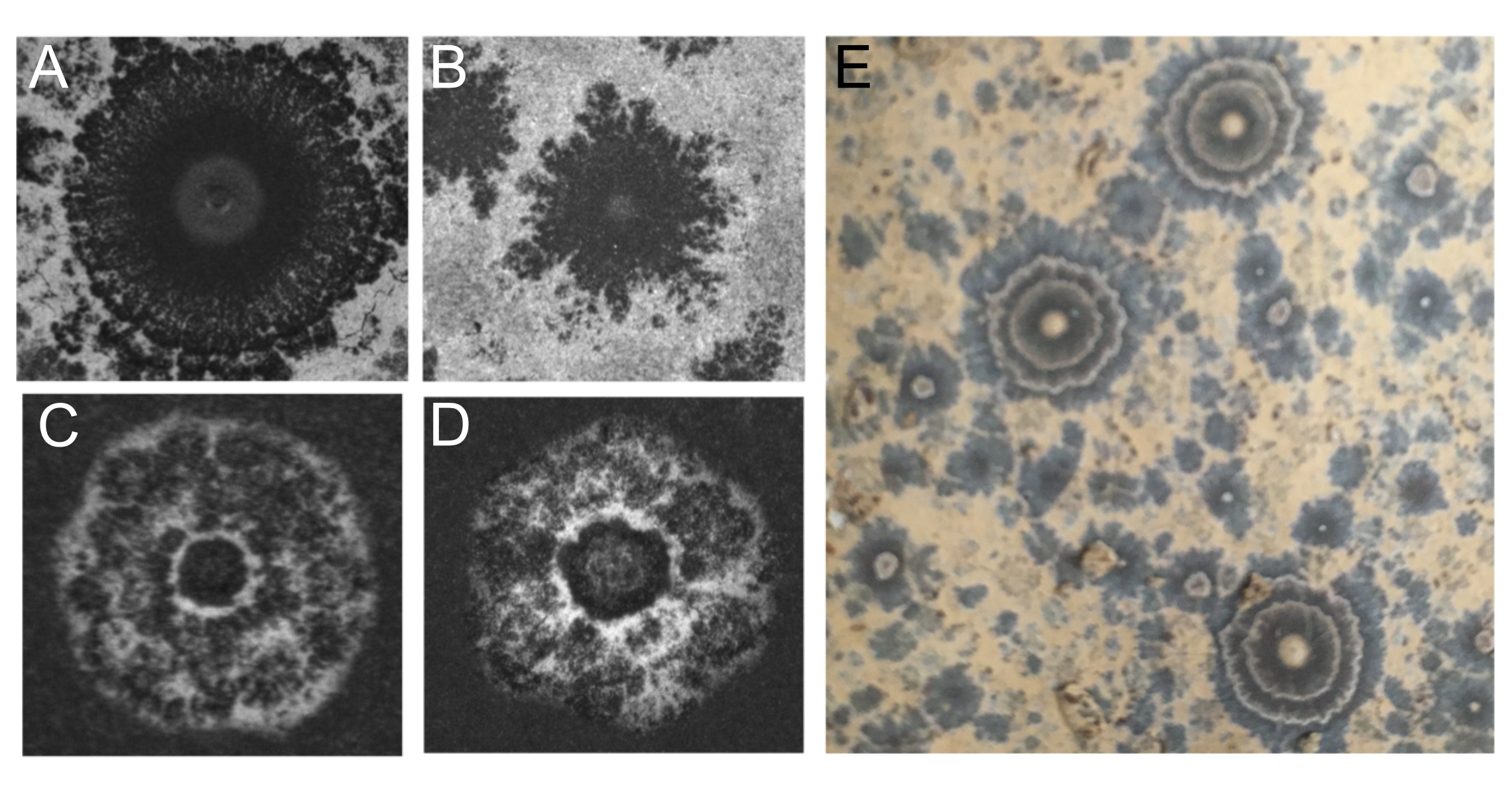}
\end{center}
\caption{Spontaneous patterns on chocolate surfaces obtained at ambient temperature after several days. A to D pictures are 5.15 mm width.}
\label{fig:patterns1}
\end{figure}

The study of chocolate is interesting for many reasons. Despite its evident importance in the food industry, this complex mixture is also a good example of some scientific curiosities \citep{Blanco2019}. For sure, some of these curiosities were found because of its technological importance and this is the case of our research: if a chocolate bar on a lab table had not become aged by chance, the observation of beautiful patterns on its surface had not inspired this ongoing study. Most of us have seen the appearance of patterns on a chocolate bar when left unperturbed for a long enough period of time, allowing it to age, some examples of such patterns are shown in Fig. \ref{fig:patterns1}. It is easy to see that the contours formed call for the use of fractals to explain them, but shape generalities might not be easy to describe in scientific terms. On a close inspection of patterns in Fig. \ref{fig:patterns1}, in the same sample, one finds a variety of shapes  as different as “mostly circular”, “snowflake-like”, “almost circular” and “hexagonal-like”.

\begin{figure}[htb]
\begin{center}
\includegraphics[width=0.55\linewidth]{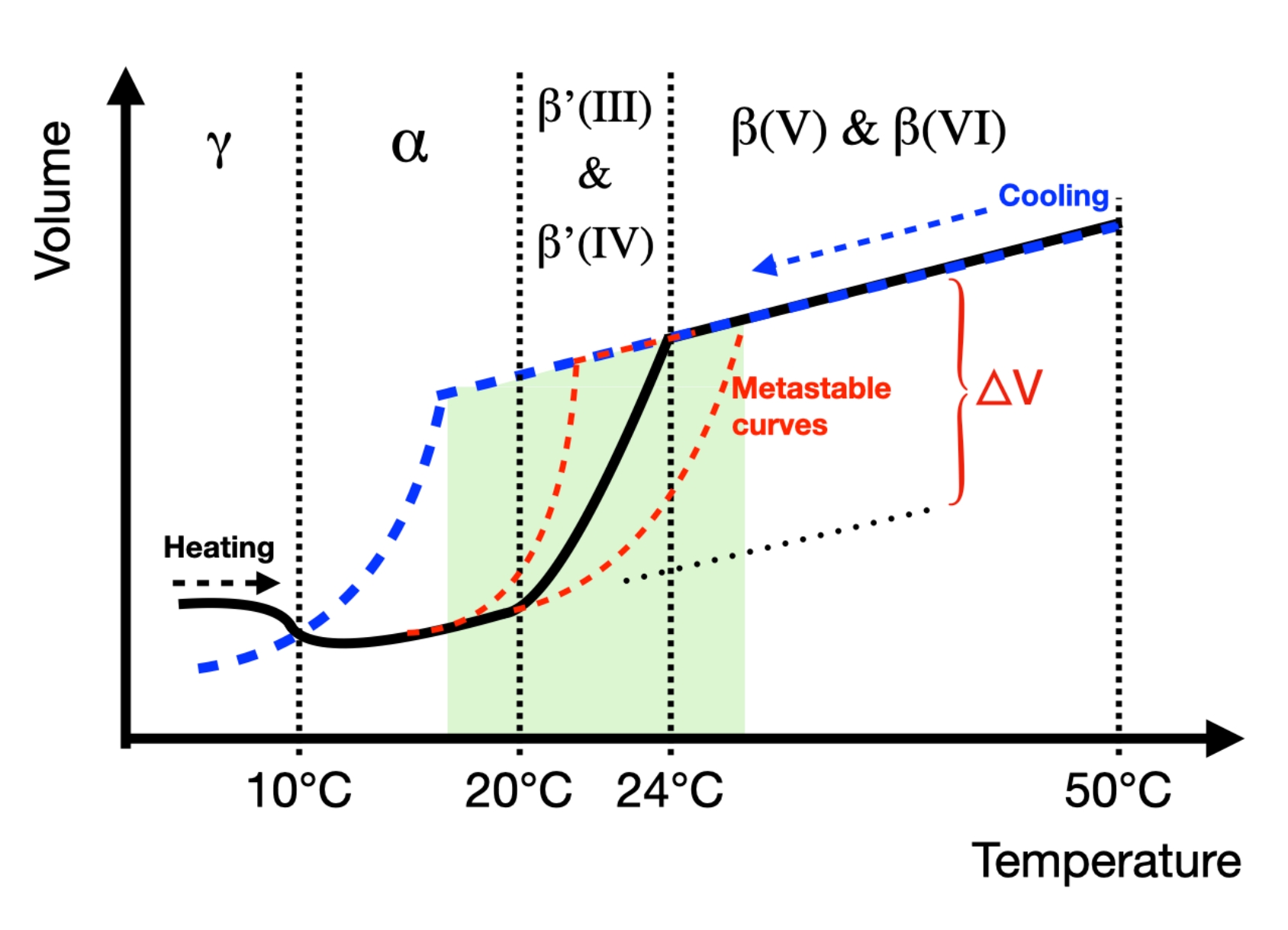}
\end{center}
\caption{Illustrative diagram of the volume change for the different phases observed under heating (black continuous line) and cooling (blue segmented line) ramps for cocoa butter. The phases more commonly described and named in the literature are annotated at the top of the different sections separated by vertical segmented lines. The intervals marked for each phase are approximated. The green region marks a common range of dispersion data associated to different phase transition and glass transition temperature values reported in literature. Metastable curves and melt expansion volume are suggested accordingly.
}
\label{fig:phases}
\end{figure}

The formation of chocolate phases and their properties can be revised extensively in the literature \citep{Sato2001}. Fig. \ref{fig:phases} summarizes the phases observed as function of temperature \citep{Marangoni2003, Gouveia2019}. In the temperature range of 50 to 60$^\circ$, all the history of the sample can be erased. The volume change in the cooling–heating processes presents an important hysteresis. On one side, under cooling it is understood that the polymorphic forms $\beta(\mathrm{V})$ and $\beta(\mathrm{VI})$ form low temperature phases without a clear formation of the phases $\alpha$ and $\beta'$. In contrast, under heating, volume changes are noticeable when different phases are formed. As depicted in Fig. \ref{fig:phases}, previous works suggest that volume changes are not monotonous \citep{Gouveia2019}. Phase change temperatures are approximate and, in general, temperatures reported differ by several degrees. Moreover, glass transition temperatures are also reported in the same ranges as the ones for phase change temperatures \citep{Calva-Estrada2020}. As a consequence, it is easy to suggest metastable lines that practically span all over the volume temperature diagram due to the proximity between the phase change temperatures \citep{VanKrevelen2009,Simha1962}.

The story is a little different for the formation of patterns on the surface of chocolate, the phenomenon is not clearly understood, and few studies analyze the shape of these patterns \citep{Marangoni2007}. In addition, from the collection of patterns previously reported, it is difficult to draw conclusions about the shapes formed, or about the reasons that gave rise to them. However, it is a consensus to consider the onset of the whitish color on the chocolate surface a consequence of the “fat bloom” event: the physical separation between $\beta(\mathrm{V})$ and $\beta(\mathrm{VI})$ phases during the aging process of chocolate. In the literature, one can find some “recipes” to produce “fat bloom”, most of them involve the use of temperature cycles \citep{Marangoni2003, Sonwai2010, RodriguezFurlan2017}. However, in our initial attempts to create such patterns, we had difficulties obtaining them using periodic cycles or constant temperatures. As a consequence, the patterns observed in Fig. \ref{fig:patterns1} were produced using a different procedure to what is usually reported in the literature.

In this paper, we reproduce the common patterns observed during the aging of chocolate and give a glimpse on the reasons for obtaining them, based on a theoretical description given by the Avrami model for nucleation \citep{Avrami1,Avrami2}. For our study we use both experiments and simulations. Definitively, the complex mixture of at least triglycerides, polyphenols and carbohydrates of high molecular weight called “chocolate”, represents a challenge for experimental reproducibility. Despite being difficult to conduct an experiment expecting the same pattern twice, it is possible to reproduce some trends of events necessary to obtain patterns that share some common characteristics. Here, it is crucial to identify, as is going to be discussed, that different events play a role in the final shape observed.

\section{Experimental and theoretical methods}

\subsection{Sample preparation}

Two types of commercial chocolate were used in this work: a commercial 100\% cacao product from the brand Mayordomo without added sugar or soy lecithin ({https://chocolatemayordomo.com.mx}) and a chocolate from the brand New Art Xocolatl with 85\% cacao ({https://www.newartxocolatl.com}). The latter has added sugar and soy lechitin. These two extra components are usually the main difference between pure cacao and the mixture called “chocolate”. In these mixtures, soy lechitin allows emulsification of the product with water and milk while sugar provides sweetness, which is completely absent in pure cacao. Despite the differences in composition between the two chocolates used, the experimental results were similar when changing the type of chocolate used. 

Three different kinds of experiments were conducted: observation of formed patterns in time on chocolate surfaces at different temperatures, atomic force microscopy of the initial events on the pattern formation and rheology of melted chocolate. Patterns on chocolate surfaces were obtained by heating up samples of chocolate up to 60$^\circ$C and pouring the viscous liquid in pre-heated petri dishes, which were then subjected to different temperature conditions. Time and temperature were recorded during pattern formation. The images analyzed in this paper were captured from these petri dishes during the evolution of the patterns. The topography Atomic Force Microscopy (AFM) images were also obtained from the petri dishes samples using a Witec alpha-300 microscope in a tapping mode ($100 \, \mu \mathrm{m} \, \times 100  \, \mu \mathrm{m}$ and $ 150 \, \times \, 150$ lines). Rheology was performed in a HR-3 Discovery Hybrid Rheometer (TA Instruments) using a cone-plate geometry of 40 mm diameter and 0.5 deg. The temperature was controlled in the  rheometer using a Peltier plate. 

\subsection{Computer simulations}

\begin{figure}
\begin{center}
\includegraphics[width=0.45 \linewidth]{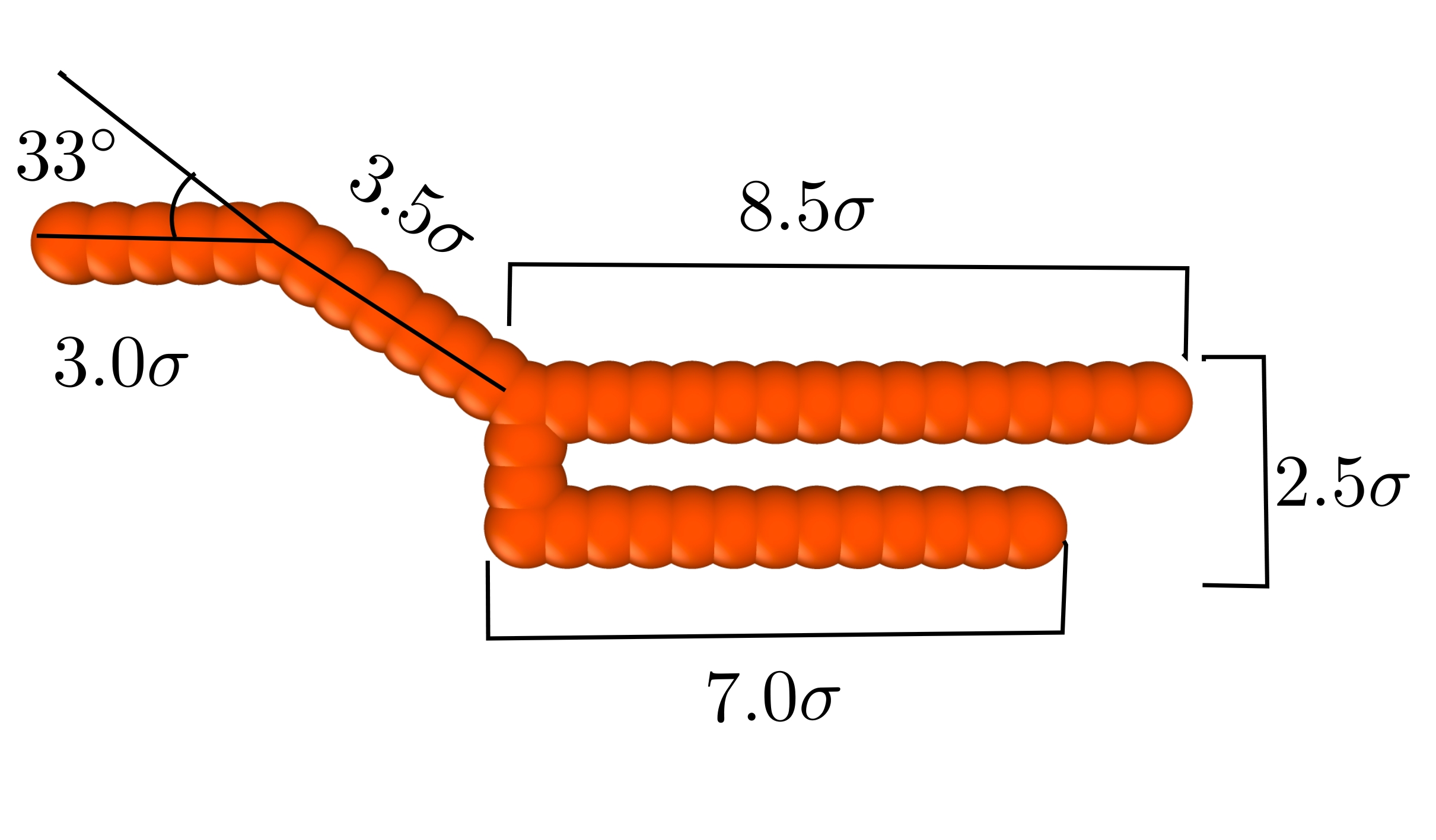}
\end{center}
\caption{Illustration of the particles used to model $\beta(\mathrm{V})$ molecules. Each particle is composed of 43 fused spheres of diameter $\sigma$ that form a chair-shaped molecule. The length of each segment of the molecule is shown in the diagram.}
\label{fig:sim_molecule}
\end{figure}

We have developed a two-dimensional (2D) model for the experimental system which allow us to have a glance at the self assembly of the $\beta(\mathrm{V})$ molecules. Our system consist of rigid bodies, each made with spheres of diameter $\sigma$, that have a shape close to one of a $\beta(\mathrm{V})$ molecule \citep{vanMechelen2006,vanMechelen2004,Hernqvist1990}, see Fig. \ref{fig:sim_molecule}. We have chosen a Mie-type potential to model the short-range attractive behaviour between fat molecules. This type of short-range interactions have been previously used to model units of fat crystals and its assembly into bigger systems \citep{Maragoni_softmatter_2012}.  The spherical units in each molecule interact only with the spheres of neighboring molecules via the pair potential,

\begin{equation} \label{eq:potential}
u_{ij}(r)= \left\{
  \begin{array}{l l}
     C \epsilon \left[ \left( \frac{\sigma}{r} \right)^{p}-\left( \frac{\sigma}{r} \right)^{q} \right]+ U_{0} & \quad r < r_{cut}\\\\
     0 & \quad r \geq r_{cut}
  \end{array}. \right.
\end{equation}
Where $r$ is the center-of-mass distance between two spheres, $ij$ indicates interaction between molecules $i$ and $j$, $\sigma$ is the approximated diameter of the repulsive core and $\epsilon$ is the strength of the interaction in $k_B T$ units. The exponents $p$ and $q$ in Eq.~\ref{eq:potential} define the range of the attractions, and are set to $p\!~=~\!36$ and $q\!~=~\!24$. The constant $C\!~=~\!{p}\left({p}/{q}\right)^{q/(p-q)}/{(p-q)}$ ensures that the minimum of the potential is at $\epsilon\!~=~\!1$. The range of the interaction between molecules is set with the cut-off parameter $r_{cut}\!~=~\!1.9\sigma$. The shift value $U_{0}$ is chosen accordingly such that $u_{ij}(r_{cut})\!~=~\!0$. For simplicity, we have set $\epsilon\!~=~\!1$, $\sigma\!~=~\!1$ and the unit time $\tau_{sim}\!~=~\!\sqrt{m\sigma^2/k_B}$.

All simulations have been performed with the open-source MD simulation package LAMMPS \citep{plimpton1995lammps}, which has a dynamical integrator for rigid bodies \citep{miller2002,kamberaj2005}. In our 2D simulations, the spherical units that form our molecules are restricted to a plane and keep the same initial $z$ position.  Each molecule segment of length $1\sigma$ is composed of two spheres, with a total of 43 spheres per molecule and 1020 molecules in our 2D simulation box. To explore the assembly of our molecules, we use the packing fractions $\phi=0.22$, $\phi=0.36$ and $\phi=0.48$. For each case, we start with a 2D box with  an initial isotropic configuration, which was created with purely repulsive molecules ($r_{cut}=\sigma$). The interaction between the molecules is turned on in a constant number, volume and temperature (NVT) ensemble using a Nose-Hoover thermostat \citep{martyna1992nose}. We have set the temperature of our systems to $k_BT/\epsilon \!~=\! 1$. The simulation time step was taken as $\delta t = 0.005\tau_{sim}$ and the total simulation times used were of $2\times 10^{5} \ \tau_\mathrm{sim}$ to $5\times 10^{5} \ \tau_\mathrm{sim}$.

\begin{figure}
\begin{center}
\includegraphics[width=0.8\linewidth]{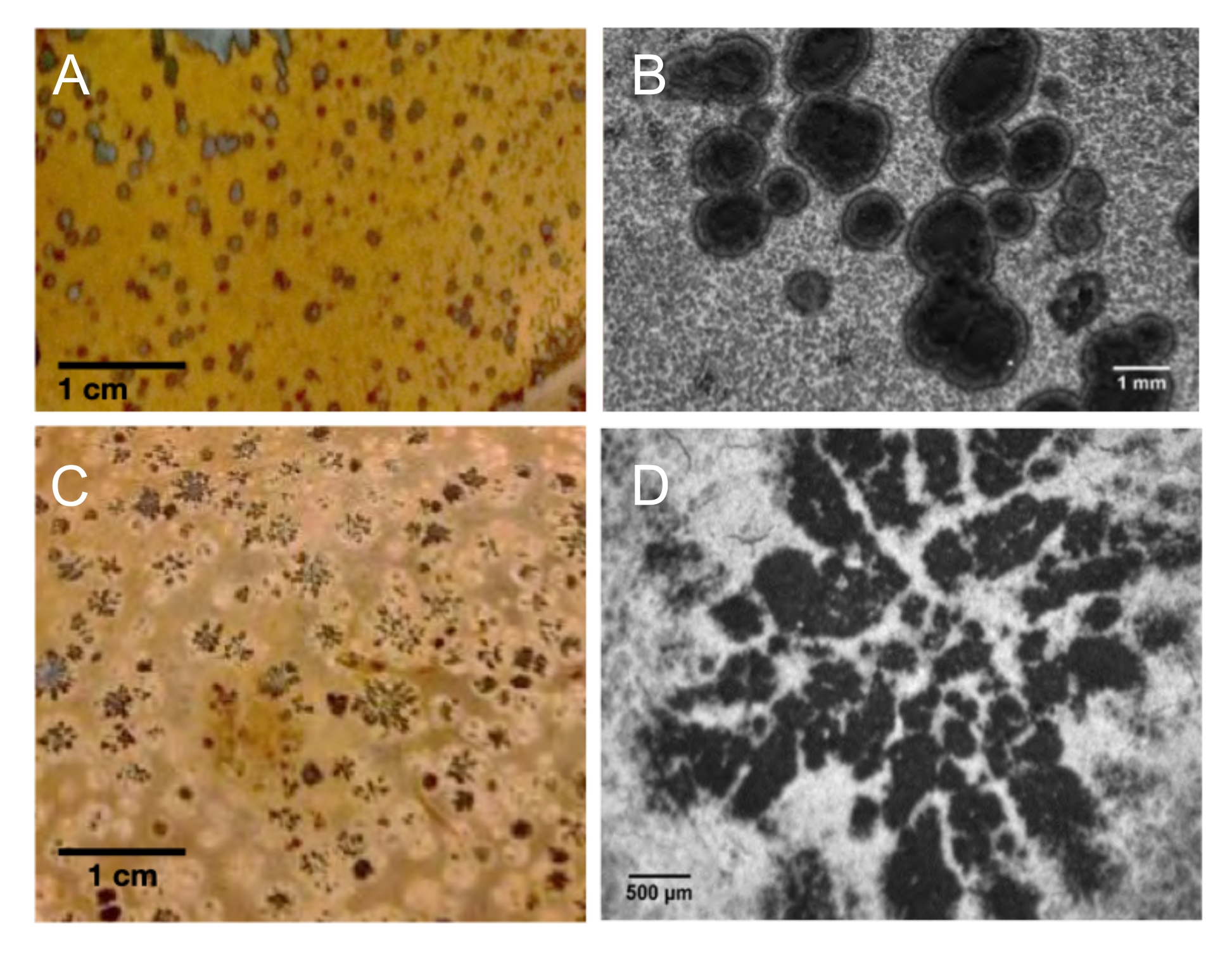}
\end{center}
\caption{Patterns on chocolate surfaces produced with a cycle or a constant temperature value. Top: patterns obtained with a temperature cycle from 17 to 35$^\circ$C in 12 hrs. during 4 days. Bottom: patterns obtained using a constant temperature of 26$^\circ$C for three days. In both cases, the samples are cooled down from 60$^\circ$C and poured on petri dishes thermalized at the minimum temperature tested.}
\label{fig:patterns2}
\end{figure}

\section{Results}

\subsection{Pattern formation}

\begin{figure}
\begin{center}
\includegraphics[width=0.8\linewidth]{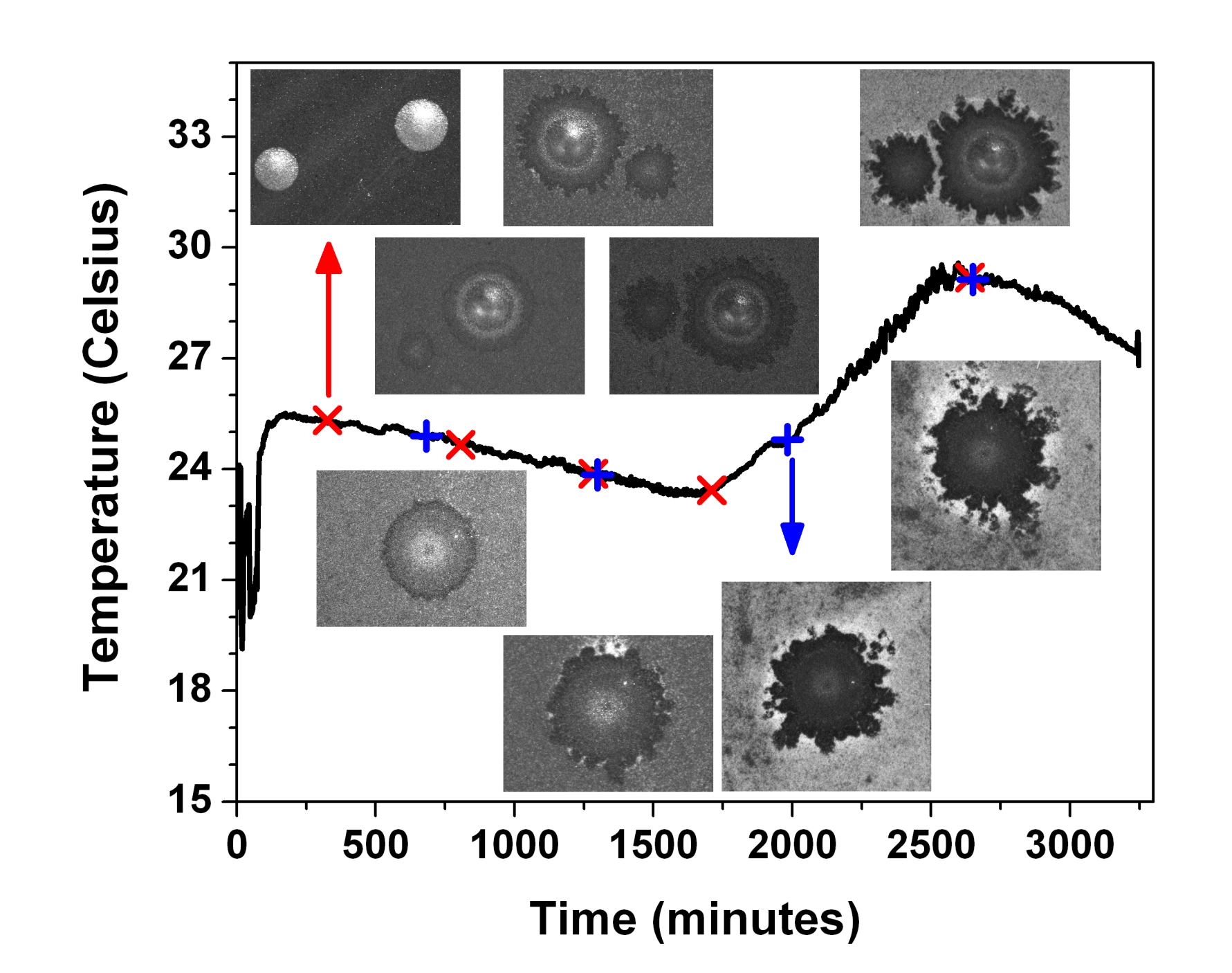}
\end{center}
\caption{Ambient temperature fluctuation in time and spontaneous patterns obtained in these conditions on a chocolate surface. Upper patterns correspond with red crosses and lower patterns with blue plus signs on the temperature curve. Patterns on the left, before 2000 min., are 5.15 mm width and the last three patterns on the right are 10.55 mm width.
}
\label{fig:temp-cycle}
\end{figure}

\begin{figure}
\begin{center}
\includegraphics[width=0.9\linewidth]{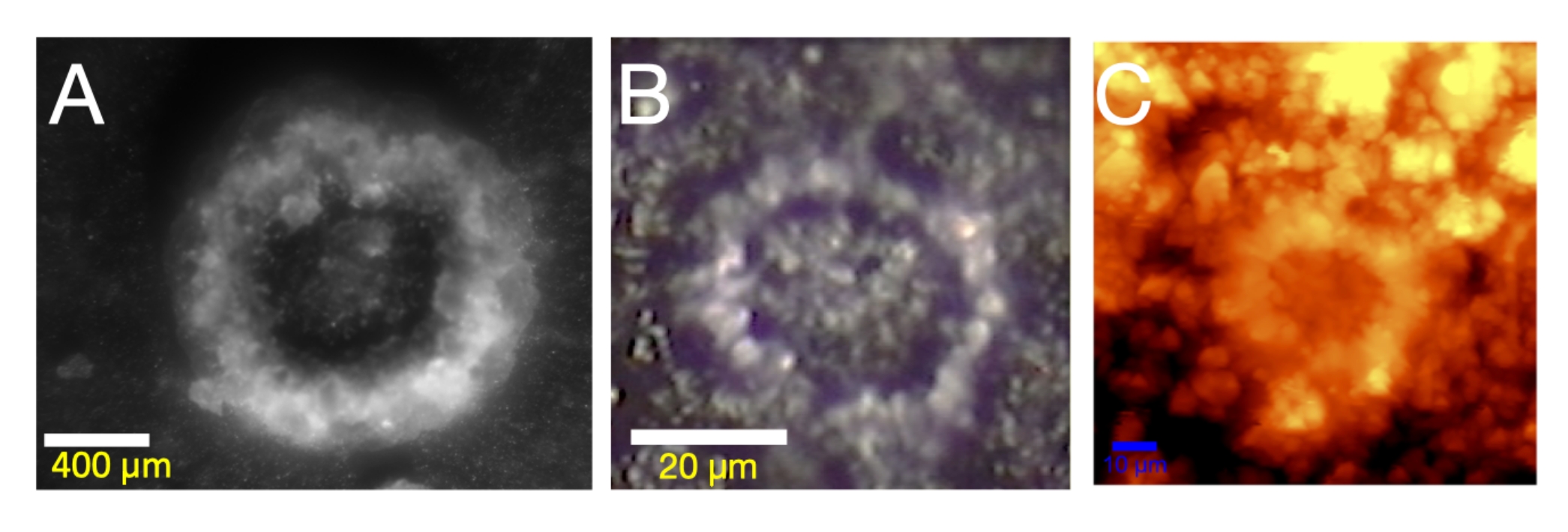}
\end{center}
\caption{(A - B) Beginning of the pattern formation at different spatial scales. (C) Topographic image of the pattern observed in (B) by AFM microscopy. Brighter zones correspond with upper regions in space.}
\label{fig:patterns3}
\end{figure}

Fig. \ref{fig:patterns2} shows several attempts to obtain the patterns observed in Fig. \ref{fig:patterns1} with different cycles of temperature, as explained in the figure caption. As it can be observed, the patterns formed by controlled cycles, or just by maintaining a constant temperature, do form different types of patterns, but they are qualitatively different from those in Fig. \ref{fig:patterns1}. Perhaps the main phenomenological differences between patterns in both figures are the occurrence of clear circular centers and more symmetric shapes, see the two patterns presented in Fig. \ref{fig:patterns1}. What we observe in Fig. \ref{fig:patterns1} is the most commonly observed phenomenology and it can only be produced when temperature fluctuates in the range between 20 to 30$^\circ$C, as observed in the sequence of pictures presented in Fig. \ref{fig:temp-cycle} with the correspondent registry of temperature. These fluctuations are, apparently, necessary to obtain the characteristic morphology  of these patterns. The initial steps to obtain this morphology are shown in Fig. \ref{fig:temp-cycle}. At the very beginning, a circular center acts as a germ nucleus for a new phase, and in this way the polymorphic forms evolve from $\beta'$ to $\beta$, as reported in the literature \citep{Gouveia2019}.  Some of the pictures suggest correctly that the center is a dome with an associated change in volume. In Fact, during all the formation of patterns, volume changes modify and produce patterns with uneven surface. The blooming is dusty, making difficult a measure with AFM, but interestingly, the beginning of the pattern can be measured with this technique as observed in Fig. \ref{fig:patterns3}. Here, we show the first steps of the pattern formation, where we see a clear circle of a different phase rising up from the surface. 

\begin{figure}
\begin{center}
\includegraphics[width=0.8\linewidth]{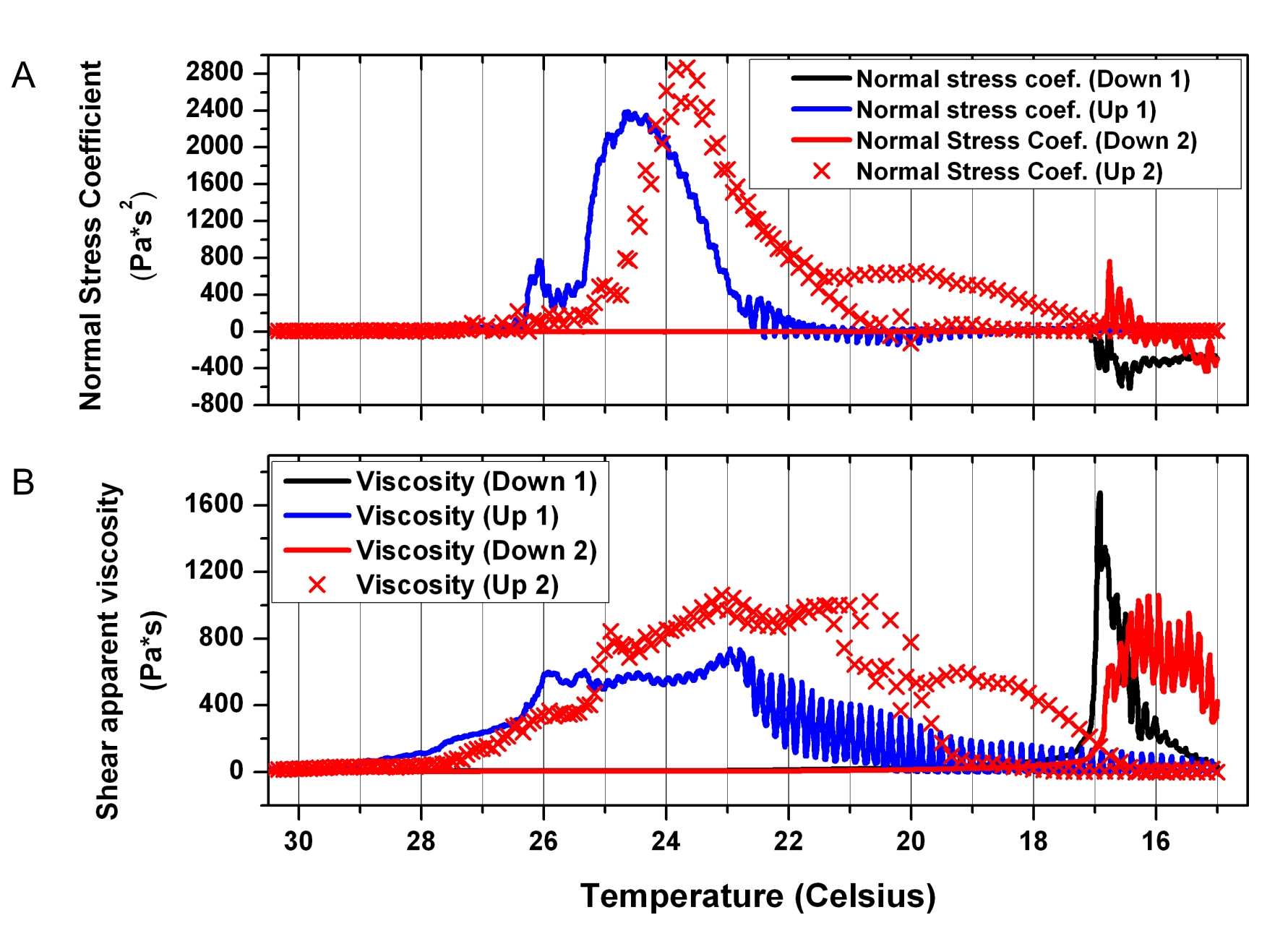}
\end{center}
\caption{ First Normal Stress Coefficient (A) and Shear Apparent Viscosity (B) for chocolate in a temperature cycle using a rate of 0.25$^\circ$C/min. The sample was first conditioned 5 min at 60$^\circ$C prior to the beginning of the experiment. Then, the sample was cooled down from 40$^\circ$C to 15$^\circ$C (“Down” in the legend) and immediately heated up again until 40$^\circ$C (“Up” in the legend). The experiment was performed twice with two different samples (“1” and “2” in the legend).}
\label{fig:shear}
\end{figure}

Using rheology, it is possible to correlate volume changes and reproducibility of the melting process. Fig. \ref{fig:shear} shows shear apparent viscosity and the first normal stress coefficient of a sample under temperature cycles. For the two chocolate types, the viscosities rise up under cooling clearly because of the formation of the melting phase. The first time that the sample was cooled, the viscosity of the samples raised up at a temperature around 18$^\circ$C, but the temperature at which the maximum occurs is not the same. Remarkably, because of the pre-history of the samples, this difference exacerbates under heating. Moreover, between two different experiments with the same cooling or heating history, it was impossible to obtain reproducible temperatures where the same viscosity rose up. Interestingly, the temperatures around which we observe the rise of both viscosities (18$^\circ$C and 25$^\circ$C) have been described as glass-transition temperatures, which is possibly related to traces of unstable polymorphic forms of monounsaturated triacylglycerides \citep{Calva-Estrada2020}. It is important to point out that, in our case, the fact that the first normal stress coefficient, related with the change in volume, could rise up in different moments even for the same conching history, might be one of the reasons for the wide range of pattern shapes observed on the chocolate surfaces.

\subsection{Simulation results}

\begin{figure}
\begin{center}
\includegraphics[width=0.5\linewidth]{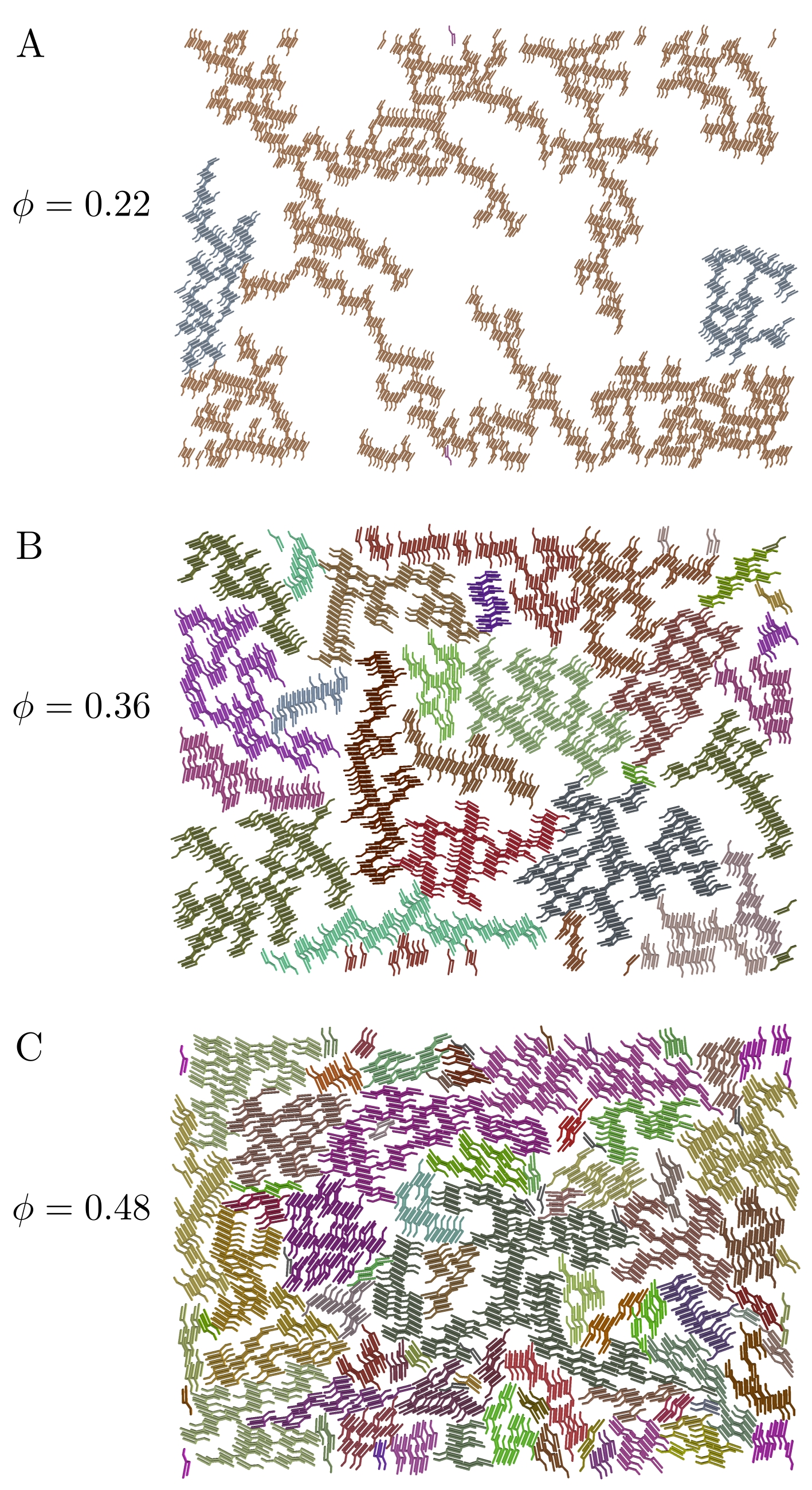}
\end{center}
\caption{Snapshots of typical final configurations of simulation boxes at packing fractions $\phi=0.22$, $\phi=0.36$ and $\phi=0.48$. Colors indicate different clusters, the prefered orientation of the molecules changes between clusters.}
\label{fig:2d_snapshots}
\end{figure}

We carried out simulations with packing fraction values $0.22 \leq \phi \leq 0.36$, in order to explore the assembly of fat molecules. This range was chosen considering the unit cell length values in crystalline polymorphic forms of triglycerides and the elongated shape of the molecules, considering that effective packing fractions around $0.30$ can be proposed for these phases. Fig. \ref{fig:2d_snapshots} shows snapshots of the typical configurations obtained for the concentrations explored. For each case, clusters with similar orientations are identified with different colors. The clusters indicate groups of molecules, with a relative angle $\theta$ of the main axis with the closest neighbors that follows $\theta \leq 10 ^\circ$. Here, our analysis of the systems is purely qualitatively, and focuses only on the cluster sizes and connectivity, but relevant information can be extracted from it. In Fig. \ref{fig:2d_snapshots}A, only two different clusters are found. When the concentration is increased, the connectivity of the molecules is reduced and several small clusters appear (see Fig. \ref{fig:2d_snapshots}C). Thus, Fig. \ref{fig:2d_snapshots} shows that the clusters become smaller as the concentration increases. This observation is relevant because these clusters correspond to the nuclei present in chocolate samples, and their size might also have an effect on the final shape of the patterns formed on chocolate surfaces.

\subsection{Pattern growth}

The nucleation process observed in experiments and simulations can be theoretically described using the Avrami model \citep{Avrami1,Avrami2}, which is based on a statistical description of the formation of grains starting from initial or germ nuclei. The basic assumption behind this model is that the factors that govern the tendency of the growth nuclei are similar to those which govern further growth, i.e, what Avrami described as an {\it isokinetic process}. 
If $p(t)$ describes the probability of formation of growth nuclei per germ nucleus per unit time at temperature $T$, then
\begin{equation}
p(t)=Ke^{-(Q+A)/RT}
\end{equation}
where $Q$ is the energy of activation per gram molecule, $A$ is the work per gram molecule required for a germ nucleus to become a growth nucleus, $R$ is the gas constant and $K$ is a normalization constant. The probability $p(t)$ is used to introduce a characteristic time $\tau$ of the growth process, defined as
\begin{equation}
    \frac{d\tau}{dt} = p(t).
\end{equation}
In a similar way, a spatial scale $r(t)$ that defines the size of the grain is given in terms of a rate of growth, $G(t)$, according to the expression
\begin{equation}
    \frac{dr}{dt}=G(t).
\end{equation}
Both equations allow us to define a volume $V$ and a surface area $S$ of a grain,
\begin{eqnarray}
    V(\tau,z)&=&\sigma_{3D}r(\tau,z)^3 \\
    &=&\sigma_{3D} \left[ \int_z^\tau\frac{G(x)}{p(x)}dx \right]^3,
\end{eqnarray}
for the three-dimensional case (3D). For a strictly two-dimensional (2D) system we have
\begin{eqnarray}
    S(\tau,z)&=&\sigma_{2D}r(\tau,z)^2 \\
    &=&\sigma_{2D} \left[ \int_z^\tau\frac{G(x)}{p(x)}dx \right]^3,
\end{eqnarray}
where $z$ is an initial value of the characteristic time, $\sigma_{3D}$ and $\sigma_{2D}$ are 3D and 2D shape factors that in the case of spherical or disk-like grains take the values $\sigma_{2D} = \pi$ and $\sigma_{3D}=4\pi/3$, respectively. The isokinetical regime consists in assuming that the rate of growth is proportional to the probability of formation, {\it i.e.}, $G(t)=\alpha p(t)$, where $\alpha$ is a constant over concentration and temperatures. This assumption simplifies the evaluation of $V(t)$ and $S(t)$, resulting in the expressions
\begin{equation}
    V(\tau,z)=\sigma_{3D}(r-z)^3
\end{equation}
and
\begin{equation}
    S(\tau,z)=\sigma_{2D}(r-z)^2
\end{equation}

\begin{figure}
\begin{center}
\includegraphics[width=0.7\linewidth]{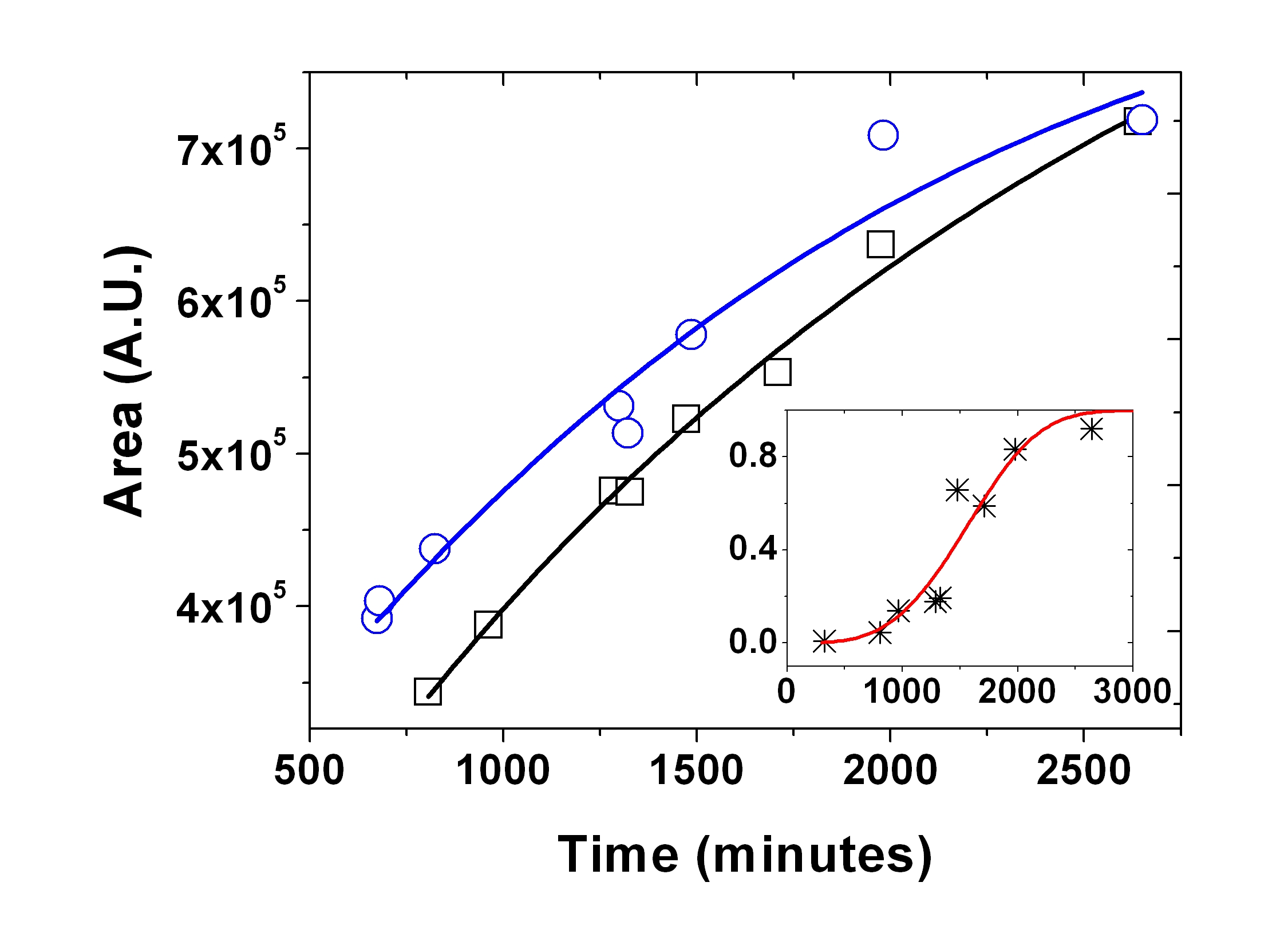}
\end{center}
\caption{Growing area vs. time for the patterns observed in the two series of Figure 4. The solid line is the fitting of the data to the 2D Avrami equation. The temporal Avrami exponent obtained is $d= 0.905 \pm 0.14$ (open square) and $d=01.142 \pm 0.28$. The square correlation coefficient is higher than 0.96 in both cases. Inset: Normalized fat bloom growing area from binarized images vs. time of a section of the chocolate surface where no patterns were formed; data were also fitted with the Avrami equation (solid line) obtaining $d=3.777 \pm 1.10$.}
\label{fig:avrami}
\end{figure}

These expressions correspond to isolated growth nuclei. By considering two associating mechanisms for the growth of a grain, via the nucleation around a germ nucleus or the nucleation around an already growth nucleus, and using the expressions obtained for the $V$ and $S$, as initial values for the grain sizes, Avrami\citep{Avrami1} derives rate equations for the number density of nuclei germs $N(t)$ and for the total extended volume of the grain. By coupling these equations and solving them in the general case of a  random distribution of initial germ nuclei, a general equation for the actual value of $V(t)$ is obtained, known in the literature as the Avrami equation, given by
\begin{equation}
    V(t)=V_0(1-e^{-kt^d})
\end{equation}
where $k$ is a coefficient that depends on $G(t)$ and $p(t)$ and $d$ is the Avrami exponent. Both quantities depend also on the geometric dimension of the grain. Typical theoretical values obtained for $d$ vary within the range $1 \leq d \leq 4$, depending on the value of the probability of formation of growth nuclei $p(t)$, which it is determined by the energy of activation $Q$ and the work required to form a growth grain $A$ \citep{Avrami2}. In the case of quasi two-dimensional, plate-like, grains, $d\approx 3$ if $p(t)$ is small ({\it i.e.}, a high total energy $E=Q+A$ per mole required for a germ nucleus to become a growth one) whereas for high values of $p(t)$ ({\it i.e}, small values of $E$), $d\approx 2$. Linear-like grains will have $1\leq k \leq 2$, although it is important to bear in mind that the key factor that determines $k$ is a complex one since not only depends on $E$ but on $G$. For fluctuations in temperature, we could expect that $Q$ becomes a determinant factor in the way that the grain growth formation is induced.  

The growing area of the two pattern series observed in Fig. \ref{fig:temp-cycle} was analyzed using the Avrami equation. Each pattern was drawn on a mask and this area in pixels was plotted vs. time. Results of the pattern growing for the two series, observed in the Fig. \ref{fig:avrami}, show a remarkable similarity. The fitting of the data to the Avrami equation gives an exponent $d\approx 1$. Furthermore, we analyzed the fat bloom homogeneous phenomena by performing a binarization that identifies regions of higher fat content, which presumably corresponds to $\beta(\mathrm{V})$-$\beta(\mathrm{VI})$ rich regions, and has a lighter typical color. We focused on sections of the images where no patterns grow and fitted these data to obtain an Avrami exponent $d\approx 3$.

\section{Discussion}

We have presented the different aspects required to obtain the most common patterns observed during the aging on the surface of the chocolate. One basic aspect is the necessity of having temperature fluctuations in the range of 20 to 30$^\circ$C, as observed in Fig. \ref{fig:temp-cycle}. Without them, it is practically impossible to obtain patterns with circular centers and high radial and angular symmetry. It is important to point out that the different phases observed with the variation of temperature have differences in volume and crystallinity (orientation). In the literature there are descriptions of typical hysteresis in volume between the cooling and the heating histories, as depicted in Fig. \ref{fig:phases}, which is similar to what it is suggested from hysteresis in the normal stress coefficient versus temperature (Fig. \ref{fig:shear}). Following Fig. \ref{fig:phases}, we can also suggest that the proximity among the different stable and metastable chocolate phases produces easily different shapes or patterns on its surface, that depend on temperature fluctuations and on the previous history of the sample. Clearly, more than two phases could be involved to create a pattern, but interestingly the pattern growing can be captured using the Avrami model (Fig. \ref{fig:avrami}). An Avrami exponent $d$ in the range of one is obtained and, as previously mentioned, it can be linked, via the probability of formation of growth nuclei $p(t)$, with a low value of energy $E$ of the fat bloom event triggered in a pattern. By contrast, the homogeneous fat bloom formation also analyzed according to the Avrami model (Fig. \ref{fig:avrami}, inset), present an exponent near 3, that corresponds to an energy higher than the one necessary to form a pattern.

Although the time and size scales involved in experiments and simulations are very different, the fact that the spontaneous pattern growth observed on the chocolate surface can be closely described by the Avrami equation suggests that the isokinetical approach intrinsic to this model, {\it i.e.}, the proportionality between the growth rate $G(t)$ and the probability of formation growth of nuclei, $p(t)$,
can be linked to the way the triacylglycerides molecules interact and start the growth nuclei observed in simulations. In other words, the associating mechanisms that can be modeled using the coarse-grained model of molecules in our NVT simulations scale accordingly to the Avrami model. Within their uncertainty values, the Avrami constants obtained for the experimental system are consistent with the model assumed, an indication that  it is possible to postulate a significant correlation between the parameters of the Mie pair-interaction used in the simulations and the actual value of the total energy $E$ required to form a growth nucleus from a germ one. 

In spite of the complexity of the behavior of real patterns on the chocolate surface, it is interesting that a coarse-grained model of the molecular behavior involved can be accomplished through the Avrami model in order to give information of the associating mechanisms that determine the actual growth process of chocolate grains. Furthermore, the alignment of the nuclei observed in the simulations resemble what is expected in the formation of fat crystals \citep{Fernandes2012,Maragoni_softmatter_2012,vanMechelen2006}. The difference in the connectivity for different packing fractions might be another crucial factor in fat bloom. Further issues that we would like to answer are how the fluctuation of temperature is correlated to the activation energy for the formation of grains, and the characterization of the rheological behavior of experimental samples in terms of the molecular model used in the simulation and its dependence on the Mie potential parameters.

\begin{acknowledgments}
We thank Jacqueline Ch\'avez for her help with some experiments. J. Delgado and A. Gil-Villegas thank SEP-CONACYT for the funding of the project No. 237425 “Materia Blanda Coloidal”.
\end{acknowledgments}

\bibliographystyle{naturemag}
\bibliography{chocolate}

\end{document}